\documentclass[twocolumn,prl,aps,amsfonts,showpacs]{revtex4}

\usepackage{graphicx}

\begin{document}
\draft
\preprint{}

\newcommand{\1}{{\bf \scriptstyle 1}\!\!{1}}
\newcommand{\I}{{\rm i}}
\newcommand{\p}{\partial}
\newcommand{\D}{^{\dagger}}
\newcommand{\bx}{{\bf x}}
\newcommand{\br}{{\bf r}}
\newcommand{\bk}{{\bf k}}
\newcommand{\bv}{{\bf v}}
\newcommand{\bp}{{\bf p}}
\newcommand{\bu}{{\bf u}}
\newcommand{\bA}{{\bf A}}
\newcommand{\bB}{{\bf B}}
\newcommand{\bE}{{\bf E}}
\newcommand{\bF}{{\bf F}}
\newcommand{\bI}{{\bf I}}
\newcommand{\bK}{{\bf K}}
\newcommand{\bL}{{\bf L}}
\newcommand{\bP}{{\bf P}}
\newcommand{\bQ}{{\bf Q}}
\newcommand{\bS}{{\bf S}}
\newcommand{\bH}{{\bf H}}
\newcommand{\balpha}{\mbox{\boldmath $\alpha$}}
\newcommand{\bsigma}{\mbox{\boldmath $\sigma$}}
\newcommand{\bSigma}{\mbox{\boldmath $\Sigma$}}
\newcommand{\bOmega}{\mbox{\boldmath $\Omega$}}
\newcommand{\bpi}{\mbox{\boldmath $\pi$}}
\newcommand{\bphi}{\mbox{\boldmath $\phi$}}
\newcommand{\bnabla}{\mbox{\boldmath $\nabla$}}
\newcommand{\bmu}{\mbox{\boldmath $\mu$}}
\newcommand{\bepsilon}{\mbox{\boldmath $\epsilon$}}

\newcommand{\iLambda}{{\it \Lambda}}
\newcommand{\cA}{{\cal A}}
\newcommand{\cD}{{\cal D}}
\newcommand{\cF}{{\cal F}}
\newcommand{\cL}{{\cal L}}
\newcommand{\cH}{{\cal H}}
\newcommand{\cI}{{\cal I}}
\newcommand{\cM}{{\cal M}}
\newcommand{\cO}{{\cal O}}
\newcommand{\cR}{{\cal R}}
\newcommand{\cU}{{\cal U}}
\newcommand{\cT}{{\cal T}}

\newcommand{\be}{\begin{equation}}
\newcommand{\ee}{\end{equation}}
\newcommand{\bea}{\begin{eqnarray}}
\newcommand{\eea}{\end{eqnarray}}
\newcommand{\beqa}{\begin{eqnarray*}}
\newcommand{\eeqa}{\end{eqnarray*}}
\newcommand{\nn}{\nonumber}
\newcommand{\DD}{\displaystyle}

\newcommand{\ba}{\left[\begin{array}{c}}
\newcommand{\baa}{\left[\begin{array}{cc}}
\newcommand{\baaa}{\left[\begin{array}{ccc}}
\newcommand{\baaaa}{\left[\begin{array}{cccc}}
\newcommand{\ea}{\end{array}\right]}

\title{Teleportation of electronic many-qubit states via single photons}

\author{Michael N.~Leuenberger and Michael E. Flatt\'e
}
\affiliation{Department of Physics and Astronomy, University of Iowa,
IATL, Iowa City, IA 52242, USA}
\author{D.~D.~Awschalom
}
\affiliation{Department of Physics, University of California, Santa Barbara,
CA 93106-9530, USA}

\date{\today}

\begin{abstract}
We propose a teleportation scheme that relies only on single-photon measurements and Faraday rotation, for teleportation of many-qubit entangled states stored in the electron spins of a quantum dot system. The interaction between a photon and the two electron spins, via Faraday rotation in microcavities, establishes Greenberger-Horne-Zeilinger entanglement in the spin-photon-spin system. The appropriate single-qubit measurements, and the communication of two classical bits, produce teleportation. This scheme provides the essential link between spintronic and photonic quantum information devices by permitting quantum information to be exchanged between them.  
\end{abstract}


\maketitle

The information contained in a quantum two-level system cannot be fully copied. This limitation, which is a special case of the no-cloning theorem\cite{Wootters}, describes a fundamental difference between classical and quantum information. For a quantum information processor the operation that can partially replace copying is the ability to transfer quantum information from one system to another. When the transfer is separated into a channel of classical information and one of Einstein-Podolsky-Rosen (EPR)\cite{Einstein} correlations it is called ``quantum teleportation". 
One of the most spectacular achievements in the manipulation of quantum information\cite{Bennett2000} is the successful teleportation of the quantum superposition of a single-photon state\cite{Bouwmeester} by means of entangled photon pairs\cite{Bennett1993}.
EPR teleportation of a single photonic qubit, as originally introduced 
theoretically\cite{Bennett1993}, and realized experimentally\cite{Bouwmeester,Marcikic} requires both the generation of one maximally-entangled two-photon pair (a ``Bell state") and a two-photon entangled measurement of one member of this pair along with the original photonic qubit (a ``Bell measurement")\cite{Braunstein}. The original teleportation scheme only permitted the transfer of a single-qubit state, but recently, this scheme has been extended theoretically to the case of single-qudit states\cite{Roa} (the higher-dimensional version of single-qubit states), two-qubit states\cite{Lee}, and three-qubit states\cite{Fang}. It has also recently been shown experimentally that the entanglement between two qubits can be teleported, again by means of Bell measurements\cite{Jennewein}. Teleportation of single ionic spin states\cite{Riebe,Barrett} has also been demonstrated recently, and the Bell measurements were implemented by a Raman phase gate applied to two ions, followed by single-ion spin measurements.

Here we propose a teleportation scheme for an arbitrary number of electronic
qubits that does not require an intermediate electronic qubit, or the use of external lasers to implement a phase gate. Instead of generating and measuring Bell states between the electronic qubits, this scheme 
relies on entangling both of the qubits with a single photon, yielding  
three-particle entanglement (a Greenberger-Horne-Zeilinger (GHZ) state\cite{GHZ}) 
of the qubit-photon-qubit Hilbert space. Any qubit that can be 
entangled with a photon can be used, but for specificity we consider 
here qubits encoded in the electron spin of individual quantum dots.
Fig.~\ref{GHZTeleportation} shows our teleportation scheme
for teleporting a many-qubit state from $D$ to $D'$.  In the specific approach we describe, the establishment of spin-photon entanglement occurs naturally through conditional Faraday rotation in a microcavity.    We 
emphasize that the entanglement of the destination qubit and the photon 
can be performed first, and this photon can be retained at the origin 
indefinitely before it is entangled with the origin qubit, thus our 
procedure is teleportation, not quantum transmission. The sending of the photon from 
the destination to the origin, after it is entangled with the 
destination qubit, is the step corresponding to the distribution of EPR 
pairs in teleportation. Our EPR pairs, however, are a hybrid consisting of an entangled electronic spin and  photon polarization. We find teleportation can be implemented using only single-photon 
measurements: measurement of the polarization of the photon entangled with both qubits, and 
measurement of the spin orientation of the origin qubit via a single 
photon.

We describe below in detail our teleportation scheme for one qubit 
(see Fig.~\ref{Teleportation_singledot_ellipse}).  We consider one excess electron in one quantum dot in a general single-spin state $\left| {\psi _{\rm{e}}^{(1)} } \right>  = \alpha \left|  \uparrow  \right>  + \beta \left|  \downarrow  \right> $, where the quantization axis is the $z$ axis. In order to distinguish clearly each step of the teleportation, we introduce the times $t_A<t_A+T<t_B<t_C<t_D$, where $t_A=0$. The photon propagating in the $-z$ direction is initially linearly polarized in the $x$ direction, and interacts first with the destination spin, which is initialized parallel to $x$. Thus the destination spin-photon wavefunction  is 
$\left|\psi_{\rm{pe'}}^{(1)}(0)\right> =\left|\leftrightarrow\right>\left|\leftarrow'\right>$. The photon can virtually create an electron and a heavy hole, or virtually create an electron and a light hole (see Fig. 3). No scattering from left to right circular polarization is possible, for if  light of polarization
$\sigma_{(z)}^+$ ($\sigma_{(z)}^-$) is absorbed, Pauli blocking in the dot forces $\sigma_{(z)}^+$ ($\sigma_{(z)}^-$) to be re-emitted.  These virtual processes lead to conditional Faraday rotation, that is, conditional phase shifts $e^{iS_0^{{\rm{hh}}}} $ or  $e^{iS_0^{{\rm{lh}}}}$ of the components of the electron-photon state depending on the photon polarization and spin orientation.
After the interaction of the initially unentangled photon with the quantum dot, the resulting electron-photon state is
\be
\left|\psi_{\rm{pe'}}^{\rm{(1)}}(T)\right> = e^{iS_0^{\rm{hh}}}\left|\psi_{\rm{hh}}^{\rm{(1)}}\right> 
+ e^{iS_0^{\rm{lh}}}\left|\psi_{\rm{lh}}^{\rm{(1)}}\right>,
\label{scattered}
\ee
where $\left|\psi_{\rm{hh}}^{\rm{(1)}}\right> = \left(\left|\sigma_{(z)}^+\right>\left|\uparrow'\right> 
+ \left|\sigma _{(z)}^-\right>\left|\downarrow'\right>\right)/2$ originates from the virtual process where a photon creates an electron and a heavy hole, 
and $\left|\psi_{\rm{lh}}^{\rm{(1)}}\right> = 
\left(\left|\sigma_{(z)}^-\right>\left|\uparrow'\right> + \left|\sigma_{(z)}^+\right>\left|\downarrow'\right>\right)/2$ 
originates from the virtual process where the photon creates an electron and a light hole. Both $\left|\psi_{\rm{hh}}^{\rm{(1)}}\right>$ and $\left|\psi_{\rm{lh}}^{\rm{(1)}}\right>$ are EPR states. We define now the photon state $\left| \varphi  \right>  = \cos \varphi \left|  \leftrightarrow  \right>  + \sin \varphi \left|  \updownarrow  \right>$ with a linear polarization rotated by $\varphi$ around the $z$ axis with respect to the state $\left|\leftrightarrow\right>$ of linear polarization in the $x$ direction. We can also write $\left|\varphi\right>=\left({e^{-i\varphi}\left|{\sigma_{(z)}^ +}\right>+e^{i\varphi}\left|{\sigma_{(z)}^-}\right>}\right)/\sqrt 2$. Consequently,
\bea
\left|\psi_{\rm{pe'}}^{\rm{(1)}}(T)\right> & = & \left(e^{iS_0^{\rm{hh}}}\left|\sigma_{(z)}^+\right>+e^{iS_0^{\rm{lh}}}\left|\sigma_{(z)}^-\right>\right)\left|\uparrow'\right>/2 \nn\\
& & + \left(e^{iS_0^{\rm{lh}}}\left|\sigma_{(z)}^+\right> + e^{iS_0^{\rm{hh}}}\left|\sigma_{(z)}^-\right>\right)\left|\downarrow'\right>/2, \\
& = & \frac{e^{i\left(S_0^{\rm{hh}}+S_0^{\rm{lh}}\right)/2}}{\sqrt 2}\left(\left|-S_0/2\right>\left|\uparrow'\right> + \left|+S_0/2\right>\left|\downarrow'\right>\right), \nn
\eea
where $S_0 = S_0^{\rm{hh}}-S_0^{\rm{lh}}$. Thus the spin-photon interaction produces a conditional  single-photon Faraday rotation around the $z$ axis by the angle $\pm S_0 /2$. If $S_0=\pi/2$, the linear polarization of the incoming photon is rotated $-\pi/4$ by the spin up component, and at the same time is rotated $+\pi/4$ by the spin down component, yielding two orthogonal photon polarizations. Thus 
$\left|\psi_{\rm{pe'}}^{(1)}(T)\right>=\left(\left|\searrow\hspace{-0.35cm}\nwarrow\right>\left|\uparrow'\right> 
+ \left|\nearrow\hspace{-0.35cm}\swarrow\right>\left|\downarrow'\right>\right)/\sqrt 2$, which is maximally entangled. In order to enhance the spin-photon interaction sufficiently to achieve $S_0 = \pi/2$, each quantum dot should be placed in its own microcavity (as shown in Fig.~2). Using a switchable cavity, as will be described below, permits the precise control of the Faraday rotation angle $S_0/2$ necessary for high fidelity teleportation.  After interacting with the spin at $D'$ the  photon is sent to $D$, and can be retained as a resource for teleportation from $D$ to $D'$ for as 
long as desired.

\begin{figure}[htb]
\includegraphics[width=8cm]{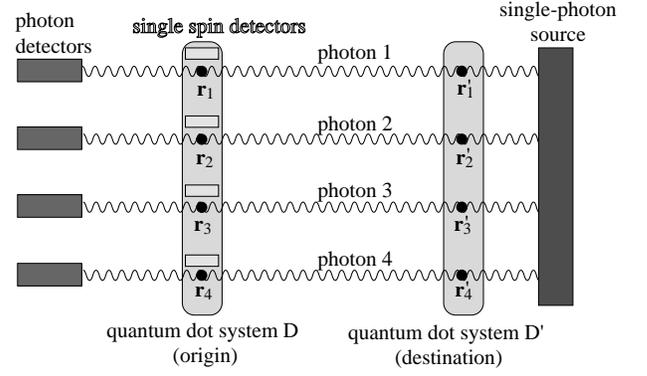}
\caption[]{Teleportation of the many-qubit state located initially in the quantum dot system $D$ to the quantum dot system $D'$. Each quantum dot of $D$ at $\br_j$ is connected to the quantum dot of $D'$ at $\br_j'$ through the photon $j$. This requirement can be satisfied using quantum dots of different sizes, where each pair of dots at $\br_j$ and $\br_j'$ have the same size. Then each pair of dots can be connected by a photon with the proper resonant frequency.  Another approach could use fiber-optics to ensure a unique connection between each pair of dots at $\br_j$ and $\br_j'$. The independent single spin detection apparatus required for teleportation is sketched in Fig. 2.}
\label{GHZTeleportation}
\end{figure}

When it is time to teleport the spin at $D$ to $D'$ we let the photon interact with the quantum dot at $D$, giving rise to a GHZ state in the hybrid spin-photon-spin system.
After this interaction we obtain
\bea
\left|\psi_{\rm{epe'}}^{\rm{(1)}}(t_C)\right> & = & \frac{e^{i\left(S_0^{{\rm{hh}}}+S_0^{{\rm{lh}}}\right)/2}}{\sqrt 2}\left(\alpha\left|\uparrow  \right>\left|-S_0/2-\pi/4\right>\left|\uparrow'\right> \right. \nn\\
& & + \alpha\left|\uparrow\right>\left|-S_0/2+\pi/4\right>\left|\downarrow'\right> \nn\\
& & + \beta\left|\downarrow\right>\left|+S_0/2-\pi/4\right>\left|\uparrow'\right> \nn\\
& & + \left.\beta \left|\downarrow\right>\left|+S_0/2+\pi/4\right>\left|\downarrow'\right>\right). 
\eea
Choosing $S_0=\pi/2$, we obtain
\bea
\left|\psi_{\rm{epe'}}^{\rm{(1)}}(t_C)\right> & = & \frac{1}{\sqrt 2}\left[\left|\updownarrow\right>
\left(-\alpha\left|\uparrow\right>\left|\uparrow'\right> + \beta\left|\downarrow\right>\left|\downarrow'\right>\right)
\right. \nn\\
& & + \left.\left|\leftrightarrow\right>\left(\alpha\left|\uparrow\right>\left|\downarrow'\right> + 
\beta\left|\downarrow\right>\left|\uparrow'\right>\right)\right], \label{updownrep}
\eea
which is 
\bea
\left|\psi_{\rm{epe'}}^{\rm{(1)}}(t_C)\right> & = & \frac{1}{\sqrt 2}\left|\updownarrow\right>\left[\left|\leftarrow\right>\left(-\alpha\left|\uparrow'\right> + \beta\left|\downarrow'\right>\right) \right.\nn\\
& & +\left. \left|\rightarrow\right>\left(-\alpha\left|\uparrow'\right> - \beta\left|\downarrow'\right>\right)\right] 
\nn\\
& & +\frac{1}{\sqrt 2}\left|\leftrightarrow\right>\left[\left|\leftarrow\right>\left(\beta\left|\uparrow'\right> 
+ \alpha \left|\downarrow'\right>\right) \right.\nn\\
& & +\left. \left|\rightarrow\right>\left(\beta\left|\uparrow'\right>  - \alpha\left|\downarrow'\right>\right)\right]\label{mixed}
\eea
in the $S_x$ representation for the spin at $D$.

We now perform measurements to complete the teleportation. If the linear polarization of the photon is measured first, then depending on the two initial spin orientations [see Eq.~(\ref{updownrep})], collapse of the wavefunction leaves the qubits at $D$ and at $D'$ in one of the four Bell states. After performing a single-spin measurement in $x$ direction of the spin at $D$ (which, as described below, can be done with a single photon),  the spin state at $D'$ is projected onto [see Eq.~(\ref{mixed})] $\left|\psi_{\rm{e1}}^{\rm{(1)}}(t_D)\right>=-\alpha\left|\uparrow'\right>+\beta\left|\downarrow'\right>$, 
$\left|\psi_{\rm{e2}}^{\rm{(1)}}(t_D)\right>=-\alpha\left|\uparrow'\right>-\beta\left|\downarrow'\right>$, 
$\left|\psi_{\rm{e3}}^{\rm{(1)}}(t_D)\right>=\beta\left|\uparrow'\right>+\alpha\left|\downarrow'\right>$, or 
$\left|\psi_{\rm{e4}}^{\rm{(1)}}(t_D)\right>=\beta\left|\uparrow'\right>-\alpha\left|\downarrow'\right>$ with equal probability. These projections correspond exactly to the states obtained in Ref.~\cite{Bennett1993}.
After communicating classically the outcome of the measurement of the linear polarization of the photon and $D$'s spin orientation along $S_x$ to $D'$,  the original spin state of $D$ can be reconstructed at $D'$ and teleportation is complete.  The same amount of classical communication (two bits) is required for our approach as was required in Ref.~\cite{Bennett1993}.  However, no intermediate electronic qubit was required, and the measurements are simply performed with single photons.

\begin{figure}[htb]
\includegraphics[width=8cm]{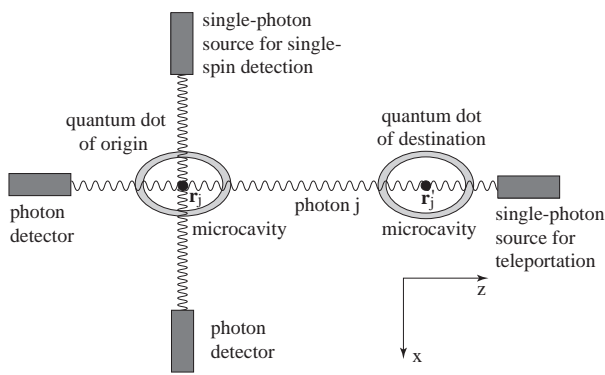}
\caption{The teleportation of the many-qubit state can be regarded as the independent teleportation of each spin state from $\br_j$ to $\br_j'$, mediated by photons traveling in the $-z$ direction. In order for the photons to scatter off the quantum dots with a probability close to unity, each dot is embedded into a microcavity. The microcavity has different lengths in the $z$ and the $x$ directions, so that the scattered photon cannot change its direction. Single-spin detection is performed by photons propagating in the $x$ direction. All the photons are measured independently from each other.}
\label{Teleportation_singledot_ellipse}
\end{figure}

The approach for two or more dots closely resembles that for one dot. The teleportation can still be performed bit-by-bit, so long as photon $j$ ($j=1,2,3,…$) coming from the dot at $\br_j'$ of $D'$ travels to the dot at $\br_j$ of $D$ (see Figs. 1 and 2). Teleportation is mediated by photons that scatter independently off the dots and the conditional phase shifts from each spin can be treated independently. This approach provides a method of teleporting a many-qubit state of an arbitrary number of qubits, always relying only on single-photon measurements.

Faraday rotation to entangle the photon and electron spin [Eq.~(\ref{scattered})] also provides the way to measure the spin with a single photon. We assume that our microcavities have an additional resonant mode at a different frequency for photons propagating in the $x$ direction. Eq.~(\ref{scattered}) shows that if the spin on the quantum dot points in the $+x$ ($-x$) direction, this incoming linearly polarized photon is converted into an outgoing circularly polarized photon 
$\sigma _{(x)}^+$ ($\sigma _{(x)}^-$). Measuring the circular polarization of the photon after it escapes yields the spin orientation along $x$.  Electrical single-spin measurements at $D$ could use instead a single electron transistor (SET), converting the spin information to charge information\cite{Loss,Friesen}.
Each of the steps along the way so far could be performed with high fidelity (time-correlated single photon counting permits a counting efficiency close to one\cite{Becker}). Accurate control of the Faraday rotation will be discussed more below.

The spin-selective coupling between the electron spins and the photons, which leads  to their mutual entanglement (and eventually to teleportation), is highly enhanced by surrounding each of the dots by its own individual high-$Q$ microcavity\cite{Gurioli} (shown in Fig. 2). Each microcavity has a single well-defined left-circularly polarized photon mode (and a right-circularly polarized photon mode of identical frequency) nearly resonant with the fundamental optical transition of the quantum dot. As we rely on nonresonant interaction of photons both in the $z$ and $x$ directions, the four highest-energy valence states should be nearly degenerate (corresponding to nearly spherical dots of zincblende or wurtzite material), i.e. their energy difference is much smaller than the detuning energy $\hbar\omega_{\rm d}$. Then, in both the $z$ and $x$ directions, we get $\sim 50\%$ selection rules (see Fig. 3), which produce a conditional phase shift (Faraday rotation) depending on the spin state of the excess electron on the quantum dot. 

\begin{figure}[htb]
\includegraphics[width=7.5cm]{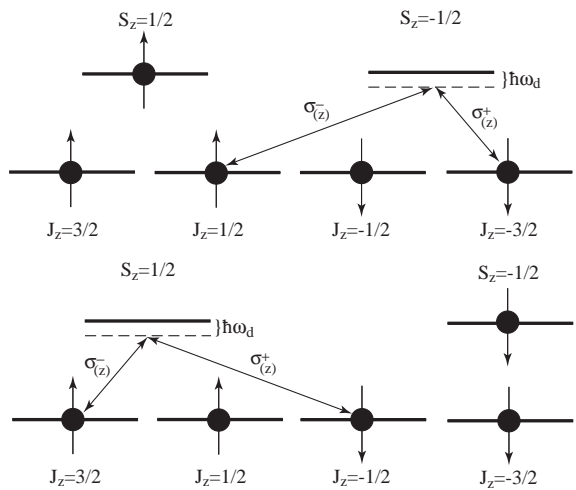}
\caption{Selection rules for the photon-dot interaction in the $z$ direction. At the top (bottom), the excess electron is in the spin up (down) state. Due to Pauli blocking, only the $\sigma_{(z)}^+$ ($\sigma_{(z)}^-$) photon can excite a virtual electron-hole pair on the quantum dot for the heavy hole states with $|J_z|=3/2$. Also, only the $\sigma_{(z)}^-$ ($\sigma_{(z)}^+$) photon can excite a virtual electron-hole pair on the quantum dot for the light hole states with $|J_z|=1/2$. Emission of the $\sigma_{(z)}^+$ ($\sigma_{(z)}^-$) photon in the $z$ direction is ensured by the boundary conditions of the microcavity shown in Fig. 2.}
\label{Selection}
\end{figure}

The Faraday rotation from the virtual process where a photon creates an electron and heavy hole is three times larger than that from the virtual process where a photon creates an electron and light hole, and is in the opposite direction\cite{Meier}. Using the electron-photon interaction energies $V_{\rm{hh}}$ and $V_{\rm{lh}}$,
$\Omega_{\rm{hh}}=V_{\rm{hh}}^2/\hbar^{\rm{2}}\omega_{\rm{d}}=3V_{\rm{lh}}^2/\hbar^{\rm{2}}\omega_{\rm{d}}$ and $\Omega_{\rm{lh}}=V_{\rm{lh}}^2/\hbar^{\rm{2}}\omega _{\rm{d}}$ are the rotation rates. Thus the phase shift accumulated by the photon state during the photon's residence time $T$ in the microcavity is given by  $S_0^{\rm{hh}}=\Omega_{\rm{hh}} T$ and $S_0^{\rm{lh}}=\Omega_{\rm{lh}} T$ for heavy and light holes, respectively. 
So the total phase shift is $S_0=S_0^{\rm{hh}}-S_0^{\rm{lh}}=\left({\Omega_{\rm{hh}}-\Omega_{\rm{lh}}}\right)T=\Omega_0 T$. If $S_0=\pi/2$ (modulo $2\pi$), the photon and electron spin become maximally entangled.

We desire the interaction strength between the photon and the quantum dot transition to be weak, i.e. $V_{\rm{hh}},V_{\rm{lh}}\ll\hbar\omega_{\rm{d}}$.
The frequency of the photon is tuned below the bandgap $E_{\rm{gap}}$, which leads to nonresonant interaction. Typical values for the bandgap and the level broadening are $E_{\rm{gap}}=1$ eV and $\Gamma=10$ $\mu$eV (see Ref.~\cite{Guest}), respectively. The interaction time between the photon and the electron spin is about $T=1$ ns (much smaller than the limiting spin decoherence time in semiconductor nanostructures\cite{Kikkawa1998}), leading to a bandwidth of 
$\Gamma_{\rm{photon}}=0.7$ $\mu$eV. If the size of the microcavity is 3.5 $\mu$m$^3$, $V_{{\rm{hh}}}$ is typically 50 $\mu$eV. Thus for a reasonable choice $\hbar\omega_{\rm{d}}\approx 1.5{\rm{ meV}} \gg \Gamma,\Gamma_{\rm{laser}}$, the scattering frequency can be adjusted to $\Omega_0=\frac{\pi}{2}\times 10^9$ s$^{-1}$, and consequently $S_0=\pi/2$.
Thus reasonable values, for a  3.5 $\mu$m$^3$ volume cavity\cite{Michler}, are $\Omega_0=\frac{\pi}{2}\times 10^9$ s$^{-1}$  and $T=1$~ns. 

To control the interaction time $T$ precisely the microcavity should be actively $Q$-switched with an electro-optic modulator. Response times of such modulators can be less than 1 ps (see, e.g. Ref. \cite{Auston}), which leads to a phase error of the order of 1 ps/1 ns = 0.1\%. The $Q$-factor can be as high as $Q = 1.25 \times 10^8$  in semiconductor structures, which is equivalent to a photon lifetime of $\tau=43$ ns (see Ref.~\cite{Armani}). Although this $Q$ was achieved for a much bigger, $10^5$ $\mu$m$^3$ cavity, there is no insurmountable reason the same processing could not be applied to cavities of the size of  3.5 $\mu$m$^3$. The theoretical limit on $Q$ for such a cavity is $\sim 10^{13}$ (see Ref. \cite{Rahachou}). Insertion of a spherical (colloidal) quantum dot into a 2D or even 3D photonic crystal with holes\cite{Srinivasan,Qi} would likely be simplest. After $T=1$ ns  in the small cavity, before $Q$-switching, the escape probability is $1-e^{-T/\tau}=2\%$. Thus the entangled state can be produced with high fidelity. 

Our teleportation scheme also provides a general link between spintronic quantum information devices and photonic ones.  Letting the photon interact only with the spin at $D$ gives the possibility to transfer $\left|\psi_{\rm{e}}(0)\right>$ onto the photon state, and back. As one example of a general class of optospintronic quantum information devices we suggest a Quantum Dynamic RAM (QDRAM) memory\cite{Bennett2000}. In such a QDRAM the many-spin state would be transferred to the many-photon state and back. As the decoherence time of photons is much longer than the decoherence time of the spins of electrons, it would be useful to keep the quantum information encoded as photons between error-correcting operations acting on the electron spins. Thus the refresh time could be much longer than the decoherence time of the electrons. Now that an efficient method of transferring quantum information between spintronic and photonic systems is available, many other such devices can be imagined which also exploit the complementary advantages of spintronic and photonic quantum information processing.

{\it Acknowledgement}. We thank Florian Meier for a critical reading of the manuscript, and Evelyn Hu for useful discussion.  We acknowledge the support of DARPA/ARO, DARPA/ONR, and the US NSF.

\end{document}